\documentclass{eas}
\usepackage{graphicx}
\usepackage{epsfig}

\begin{document}

\title{Metal Enrichment in the Intra-Cluster Medium} 
\author{S. Schindler}\address{Institute for Astro- and Particle
  Physics, University of Innsbruck, Technikerstr. 25, 6020 Innsbruck, Austria}
%
%
\begin{abstract}
The enrichment of the Intra-Cluster Medium (ICM) with heavy elements
is reviewed. There is now good observational evidence for enrichment including
abundance ratios and metallicity distributions. 
Various processes involved in the enrichment process --
ram-pressure stripping, galactic winds, galaxy-galaxy interactions,
AGN outflows and intra-cluster supernovae -- are described. Simulations
of the ICM evolution taking into account metal enrichment are presented. 
\end{abstract}
\maketitle
\section{Introduction}

The hot gas between the galaxies in a galaxy cluster -- the
intra-cluster medium (ICM) -- does not only contain primordial
elements, but also heavy elements like Fe, Si, S, or O. From X-ray
observations the overall metal abundance is known to be about $1/3$ -
$1/2$ solar (e.g. Tamura et al. 2004). Given the large mass fraction of the
ICM in a cluster (15-20\%) compared to the mass fraction of the
galaxies (3-5\%) a lot of metals must have been produced within the
galaxies and then been transported together with part of the interstellar
medium (ISM) into the ICM (Mushotzky 1999, Renzini 2004). This gas
transfer affects the evolution of galaxies and of galaxy clusters,
therefore it is important to know when and how the transport takes place.

\section{X-ray observations of metals in the ICM}

X-ray observations can also distinguish the lines of different
elements, e.g. elements like Si, S, O from core collapse supernovae or
the elements Fe and Ni from supernovae Ia
(Baumgartner et al. 2005, Ettori et al. 2002, Sanders et al. 2004,
Finoguenov et al. 2002) and can hence give information on the origin of
the metals and interaction in clusters.

Further information comes from observation of the evolution of the
metal abundance. X-ray observation can measure abundances to about a
redshift of 1. In the redshift interval from 1 to 0 they still seems
to be some evolution: an increase of the metallicity of almost a
factor of two seems possible (Balestra et al. 2006) -- but the errors are
still quite large.

Also the metallicity distribution can be measured by X-ray
observations. Azimuthally averaged metallicity profiles show a
relatively flat distribution in ``normal'' clusters, while there is an
increase in the central metallicity in cool core clusters (De Grandi
et al. 2004, Vikhlinin et al. 2005, Pratt et al. 2006). Even more
instructive are
measurements of the 2D distribution of metals -- metallicity
maps. Although it is not easy to derive them, because in each pixel
enough photons for a spectrum have to be accumulated and then fitted
with a model, many groups have derived recently quite detailed
metallicity maps (e.g. Sanders et al. 2004, Durret et al. 2005,
O`Sullivan et al. 2005, Sauvageot
et al. 2005, Werner et al. 2006, Sanders \&
Fabian 2006, Hayakawa et al. 2006, Finoguenov et al. 2006, Bagchi et
al. 2006). These
maps all show an inhomogeneous distribution of the heavy elements with
several maxima, quite different patterns and a non-spherically
symmetric distribution. The range of metallicities measured in a
cluster from minimum to maximum metallicity comprises easily a factor of two. 

\section{Enrichment processes}

Various processes can contribute to the metal enrichment. We review
here several enrichment processes, but certainly this list is not
complete. 

\subsection{Ram-pressure stripping}

One process that obtains more and more attention is ram-pressure
stripping (Gunn \& Gott 1972). There is now a lot of observational
evidence of stripped galaxies. In the Virgo cluster several
examples of ram-pressure affected spiral galaxies have been found
by HI observations (Cayatte et al. 1990; Veilleux et al. 1999,
Vollmer et al. 1999, Vollmer 2003, Kenney et al. 2004, Vollmer et
al. 2004a,b; Koopmann \& Kenney 2004, Crowl et al. 2005). Furthermore in Virgo
elliptical galaxies stripping features have been discovered (e.g.
Rangarajan et al. 1995, Lucero et al. 2005, Machacek et al. 2006a). Also in the Coma and
other clusters evidence for ram-pressure stripping has been found
(Bravo-Alfaro et al. 2000, 2001). It is even possible that the HI
plume of the galaxy NGC4388, that extends to more than 100 kpc, is
also a ram-pressure stripping feature.

As ram-pressure stripping is such a common process, there are many
models in which the stripping process of galaxies was calculated (Abadi et al.
1999, Quilis et al. 2000, Toniazzo \& Schindler 2001, Schulz \&
Struck 2001, Vollmer et al. 2001, Hidaka \& Sofue 2002, Bekki \&
Couch 2002, Otmianowska-Mazur \& Vollmer 2003, Acreman et al. 2003,
Roediger \& Hensler
2005, Roediger \& Br\"uggen 2006, Roediger et al. 2006, Mayer et al.
2006).

\subsection{Galactic winds}

Already many years ago galactic winds were suggested as a possible
gas transfer mechanism (De Young 1978). Repeated supernova explosions provide
large amounts of thermal energy, which can drive an outflow from a
galaxy (see reviews by Heckman et al. 2003 and Veilleux et
al. 2005). A correlation between starburst galaxies and wind is well
established (e.g. Dahlem et al. 1998). The outflows consist of a complex
multi-phase medium of cool, warm and hot gas. 
The morphologies of the optical
emission-line gas and the X-ray emission as observed with CHANDRA have
been studied and found to be quite
similar (Strickland et al. 2002, Cecil et al. 2002). This fact can be used
to explain the interaction between the gas 
in the bubbles and the ISM. The accelerated ISM can reach velocities of
several hundred km/s (Heckman et al. 2000, Rupke et al. 2002).

With these winds also metals are
transported into the ICM. The amount 
of metals depends on various galaxy parameters, like the total mass of
the galaxy or the disc scale length, 
and on the environmental conditions: e.g. in the centre
of massive clusters the pressure of the ICM can suppress the winds
(Kapferer et al. 2006).
Martin et al. (1999) gives an often used recipe for simulations: the winds
outflow rate is a few time the
star formations rate. Also Heckman (2003) finds
by comparing different techniques the outflow rate of the order of the
star formation rate.  Other attempts to quantify
the outflow rate include physical parameters like the 
the potential of the galaxy and cosmic rays (Breitschwerdt et al. 1991).
Also hydrodynamic simulations of outflows have been performed
(Tenorio-Tagle \& Munoz-Tunon 1998, 
Strickland \& Stevens 2000). 

In some galaxies the winds are not only driven by repeated supernova
explosions but also the AGNs are contributing to energy necessary for
the wind (see Sect. 3.4). 
Starbursts with subsequent winds can also be caused by cluster mergers
(Ferrari et al. 2003, 2005, 2006), because in such mergers the
gas is compressed and shock waves and cold fronts are produced
(Evrard 1991, Caldwell et al. 1993, Wang et al. 1997, Owen et al.
1999, Moss \& Whittle 2000, Bekki \& Couch 2003).

\subsection{Galaxy-galaxy interaction}

Another possible mechanism for removing material -- gas and stars --
from the galaxies is interaction between the galaxies (e.g.
Clemens et al. 2000, Mihos et al. 2005). While the direct
stripping effect is probably not very efficient in clusters due to
the short interaction times, the close passage of another galaxy
can trigger a star burst (Barnes \& Hernquist 1992, Moore et al.
1996, Bekki 1999), which subsequently can lead to a galactic wind
(Kapferer et al. 2005). But there can be a competing effect: the
ISM might be stripped off immediately by ram-pressure stripping
(Fujita et al. 1999, Heinz et al. 2003) and hence the star
formation rate could drop. In any case ISM would be removed from
the galaxies.

Simulations of interactions between active galaxies show a  complex interplay
between star formation and the activity of the central active galactic
nuclei (Springel et al. 2005).

\subsection{AGN outflows}

Two types of outflows from AGNs are discussed -- jets and winds-like outflows. 
There is a lot of observational evidence that jets from AGN
interacting with the ICM -- not only radio jets but also cavities in
the ICM found in X-rays (e.g.  Blanton et al. 2001, McNamara et al.
2001, Schindler et al. 2001, Heinz et al. 2002, Choi et al. 2004,
Fabian et al. 2006), in
which the pressure of the relativistic particles of the jet has pushed away
the ICM. The jets consisting of relativistic particles can entrain
some of the surrounding gas (De Young 1986).   

As the jet-ICM interaction can have an effect both on the
energetics and the metal enrichment of the ICM several groups have
started to calculate this process. Many simulations for the
energy transfer have been performed (Zhang 1999, Churazov et
al. 2001, Nulsen et al. 2002, Krause \& Camenzind 2003, Heinz
2003, Beall et al. 2004, 2006, Dalla Vecchia et al. 2004, Zanni et
al. 2005, Sijacki \& Springel 2006, Heinz et al. 2006) while only few
have attempted to calculate the metal enrichment due to entrainment by
jets (Heath et al. 2006, Moll et al. 2006).

Also for wind-like outflows there is some observational
evidence. Blue-shifted absorption lines have been observed in UV and
X-rays (Crenshaw et al. 2003). Also from X-ray imaging evidence for
nuclear outflows have been found (Machacek et al. 2006b).
There are hints for a high
metallicity of a few times solar (Hamann et al. 2001, Hasinger et
al. 2002), for high velocities (Chartas et al. 2002, 2003, Pounds et
al. 2003a,b, Reeves et al. 2003) and for considerable mass outflow
rates (Crenshaw et al. 2003, Veilleux et al. 2005).

\subsection{Intra-cluster supernovae}

There is more and more evidence for a population of stars between
the galaxies in a cluster (Gerhard et al. 2002, 2005, Cortese 2004,
Ryan-Weber et al. 2004, Adami et al. 2004). When these stars
explode as supernovae (mainly type Ia) they can enrich the ICM very efficiently because
there is no ISM pressure around them to confine the metals
(Domainko et al. 2004; Zaritzsky et al. 2004;  Lin \& Mohr 2004).
Therefore this population of stars should also be considered for
the enrichment processes in the ICM.

\section{Simulations of ICM enrichment}

Obviously the enrichment of the ICM is a complex process, which is the
result of many different mechanisms. Moreover all these mechanism do not act
separately, but there is a lot of interplay between them. Furthermore
the scales involved range from 
the sizes  of the central regions of AGN to  the sizes of galaxy
cluster.  It is not possible to cover all these scales with a simple
simulation. Therefore many groups have developed very different ways
to model this complex problem. The questions that are addressed by
these simulations concern the amount of metal, the distribution of the
metals, the time evolution of the enrichment processes and the ratio
of different elements.

\begin{figure}    
\begin{center}
\hbox{
\psfig{file=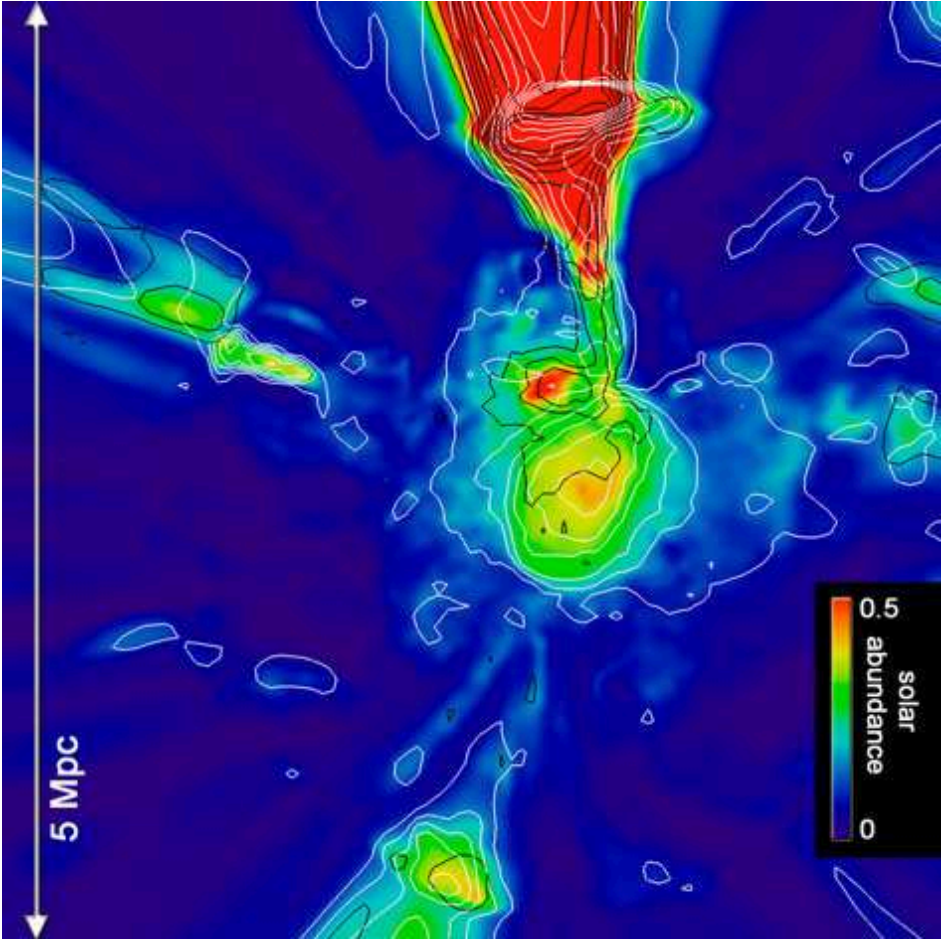,width=12.5cm,clip=}}
\caption{Simulated metallicity map = X-ray emission weighted, projected
 metal distribution of
 a merger cluster at z=0. The
  distribution is quite inhomogeneous with stripes in the outer
  parts. The high metallicity region 
  at the top is caused by a group of galaxies with recent
  starburst. Overlaid are contours which indicate the origin of the
  metals: ram-pressure stripping (white) and winds (black). 
}
\end{center}
\end{figure}

Some groups use a semi-analytical approach on top of N-body simulations,
e.g. they do not follow the motion of the ICM but are able to follow
the enrichment not only in clusters but also in the WHIM (warm-hot
intergalactic medium, De Lucia et al. 2004, Nagashima et
al. 2005, Bertone et al. 2005). They find that mainly the massive galaxies contribute to the
enrichment and that there is a mild metal evolution since z=1. Other
groups perform models without dynamics to estimate the effect of
different initial mass functions (IMF) on the ICM enrichment (Moretti et al. 2003) and the
contributions of different supernova types (Pipino et al. 2002).

\begin{figure}    
\begin{center}
\hbox{
\psfig{file=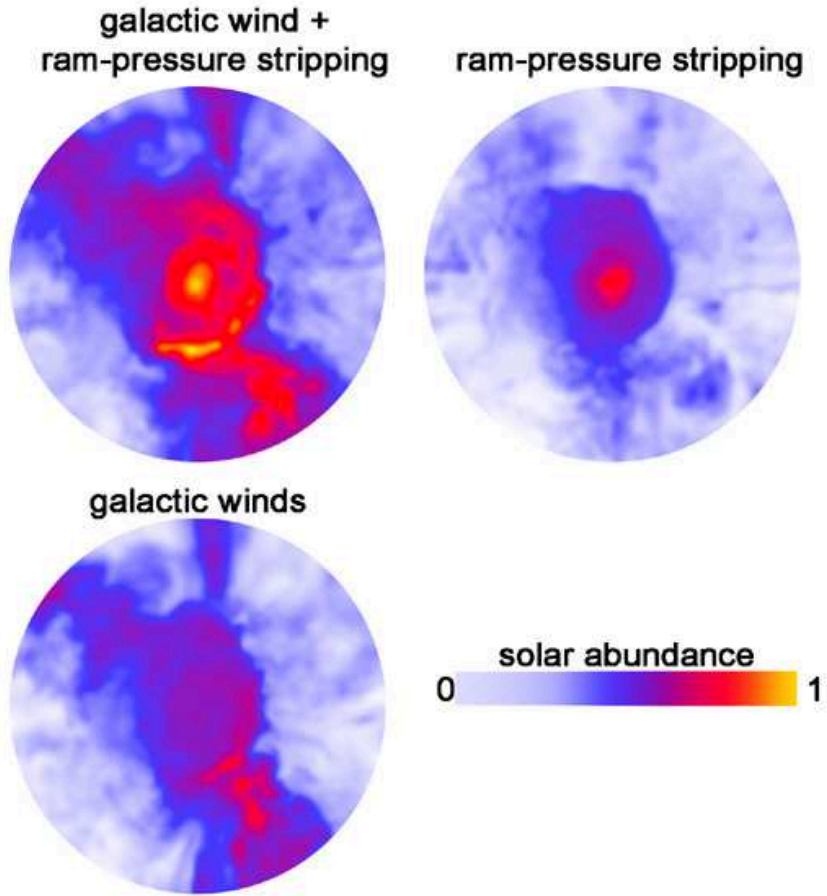,width=12.5cm,clip=}}
\caption{Comparison of the metal enrichment by ram-pressure stripping
  and winds in a galaxy cluster at z=0. The figure shows simulated
  metallicity maps with a diameter of 2.5 Mpc. Ram-pressure stripping
  yields a more centrally concentrated metallicity distribution.
}
\end{center}
\end{figure}

In a different method also the dynamics of the ICM is simulated. Some
groups use an approach with N-body/SPH simulations combined with
semi-analytical models that include detailed yields from type Ia and II
supernovae (Valdarnini 2003, Tornatore et al. 2004, Scannapieco et
al. 2005, Cora 2006, Romeo et
al. 2006). They model radial
profiles of different elements, finding relatively steep profiles
compared to observations. Also metallicity maps (Cora 2006), 
different IMFs (Tornatore et al. 2004, Romeo et al. 2006) and cooling
efficiencies of galactic objects (Scannapieco et al. 2005) are
investigated.

\begin{figure}    
\begin{center}
\hbox{
\psfig{file=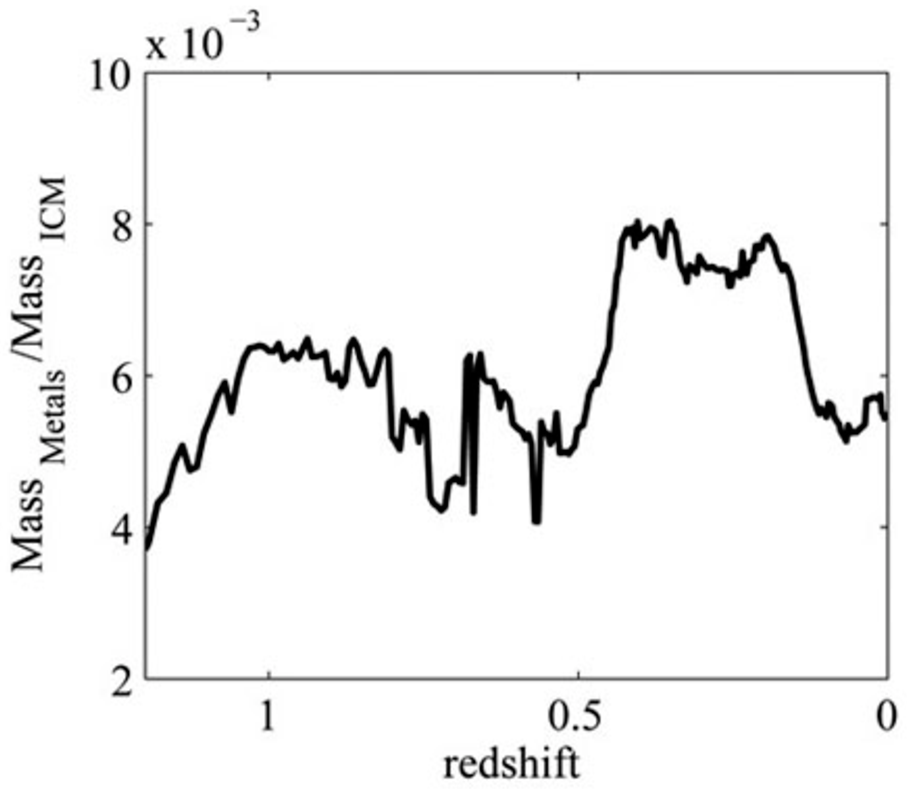,width=11.5cm,clip=}}
\hbox{
\psfig{file=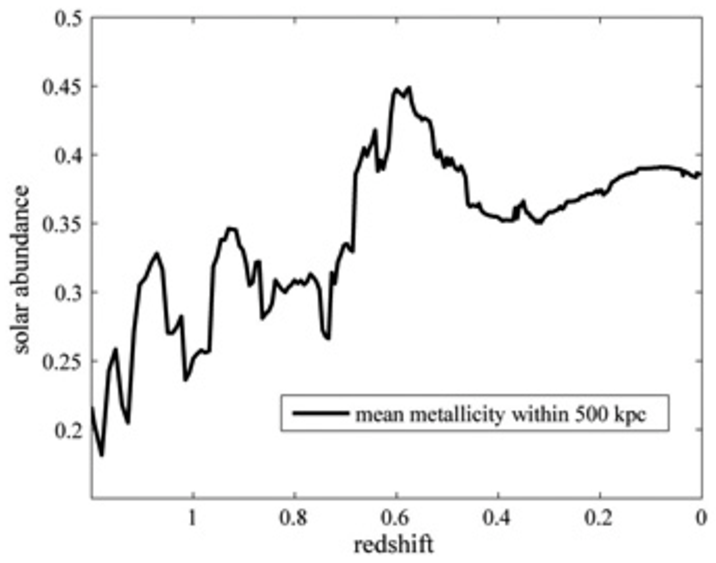,width=12cm,clip=}
}
\caption{Top: Ratio of mass of metal to the mass of the ICM within a
  spherical volume of radius 500 kpc in arbitrary units. Bottom:
  X-ray emission weighted metallicity as it would be observed if all the
  cluster emission observed within a radius of 500 kpc was used to
  derive a spectrum. Projection effects and emission weighting affect
  the observed metallicity strongly.That means a determination of the
  metal mass from the observed metallicity has very large errors. 
}
\end{center}
\end{figure}

We, the ``Hydro SKI team'', have chosen a different approach. In
N-body/hydro\-dynamic simulations with mesh refinement including a
semi-analytical method the various enrichment processes are calculated  separately with
different descriptions
(Schindler et al. 2005). So far ram-pressure stripping (Domainko et
al. 2006), quiescent and
starburst-driven winds (Kapferer et al. 2006), galaxy interactions
(Kapferer et al. 2005) and AGN outflows (Moll et al. 2006) have been
taken into account -- all of them can be switched and on and off
individually. We furthermore distinguish between ICM and
ISM, so that we can model directly virtual observations of X-ray
metallicity maps (see Fig.~1).

We find in general an inhomogeneous metallicity distribution in good
agreement with the observed metallicity maps (see Fig.~1 and
Fig.~2). Obviously the enriched material is not mixed immediately with
the ICM. Usually several processes contribute to the metallicity
of a cluster. The different processes cause different metallicity
distributions and have different time scales. In particular we
compared ram-pressure stripping and galactic winds. Ram-pressure
stripping is more efficient massive clusters, while winds can be
suppressed in the centres of massive clusters. Therefore the metals
transferred by ram-pressure stripping yield a relatively steep profile
in the centre, while the metallicity due to winds is almost
flat. The sum of the metals obtained by both processes together is in good
agreement with the observed metallicity slopes. Clusters with
sub-cluster mergers have particularly high enrichment rate due to
ram-pressure stripping.  In such clusters
ram-pressure stripping can contribute a factor of 3 more to the
enrichment than winds. Mass loss of the galaxies due to winds is stronger at early
evolutionary stages and starts to decrease between redshifts of 2 and
1. In this period ram-pressure stripping becomes stronger and
exceeds the mass loss rate due to winds. 

Metallicity values deduced from X-ray observations cannot be transferred
easily into the amount or mass of metals:
projection effects and emission weighting (depending on ICM density
and temperature) can hide/enhance the metal mass (see Fig.~3). E.g. the emission
weighted mean metallicity of a cluster is usually going up and down
with time, although there is continuously enriched material falling
into the cluster. As there is always also non-enriched material
falling into the cluster and they are not mixed immediately, there can
be quite large variations ($\approx$ 50\%) in the observed mean
metallicity. More figure and movies can be found at
http://astro.uibk.ac.at/astroneu/hydroskiteam/index.htm. 

\section{Conclusions}

Metal enrichment of the ICM is a complex process. X-ray observations
yield now a lot of information on overall metallicities, profiles, 2D
maps and abundance ratios.
Many different enrichment
processes can contribute to the metal abundance. Often the processes are not
strictly separated but influence each other. Taking all processes into
account it is not difficult to reach the observed metallicities. The
efficiencies of the different processes
vary a lot with time and with cluster environment.

\vskip 1cm\noindent
{\it Acknowledgement}  I thank the Hydro SKI Team for the very pleasurable and fruitful
  collaboration. This work was supported by the Austrian Science
  Foundation FWF (P18523), the Tiroler Wissenschaftsfonds (UNI-404/58)
  and the UniInfrastruktur 2005/2006.


\end{document}